\documentclass{article}
\usepackage{authblk}
\usepackage{tikz}
\usepackage{amsmath,mathtools}
\usepackage[round]{natbib}
\usepackage[pagebackref=true,colorlinks=true,urlcolor=blue,linkcolor=blue,citecolor=magenta]{hyperref}
\usetikzlibrary{shapes.geometric,plotmarks,arrows}
\usepackage{caption}
\usepackage{subcaption}
\usepackage{footnote}

\makesavenoteenv{tabular}
\makesavenoteenv{table}
\title{Modelling Joint Lifetimes of Couples by Using Bivariate Phase-type Distributions}
\author[1]{Hassan Zadeh A. \thanks{am\_hassanzadeh@sbu.ac.ir}}
\author[1]{Amirhashchi S. \thanks{soroush.amirhashchi@gmail.com}}
\affil[1]{Department of Acturial Science, Shahid Beheshti University, Tehran, Iran}
\begin{document}
	
\maketitle	
	
\begin{abstract}
Many insurance products and pension plans provide benefits which are related to couples,
and thus under influence of the survival status of two lives. Some studies show the future lifetime of couples is correlated. Three reasons are available to confirm this fact: (1) catastrophe events that affect both lives, (2) the impact of spousal death and (3) the long-term association due to common life style.
Dependence between lifetimes of couples could have a financial impact on insurance companies and pension plans providers.
In this paper, we use a health index called physiological age in a Markov process context by that we model aging process of joint and last survivor statuses. Under this model,  future joint lifetime of couples follows a bivariate phase-type distribution. The model has physical interpretation and closed-form expressions for actuarial quantities and owns tractable computation for the other ones. We use the model to pricing products relevant to couples annuities and life insurances.

\end{abstract}

\noindent \textbf{Keywords}: Bivariate Phase-type Distribution, Physiological Age, Markov Process, Joint Lifetimes of Couples, Aging Process, Markov processes.  

\section{Introduction and Motivation} \label{s1}
\noindent It is common for life insurance and pension products that cash flow direction depends on status of joint-life and last-survivor statuses. In joint-life insurance a lump sum is paid out on the first death and in a last-survivor life annuity the amounts of benefit is paid as long as at least one of the couples survives. In reversionary annuities the benefit is paid to one of the couples if he/she survives after his/her partner. In some pension products the pension benefit is contingent on survival of a couple and even more members of families. In this paper, we focus on modelling future lifetimes of couples rather than a group of more than two.
\par
In actuarial modelling, future lifetimes of couples are usually assumed to be independent which apparently are not.
Couples due to many reasons share risks together. Common life style, depression after bereavement of one partner and common shock are the main reasons for dependence between the future lifetimes of couples.
\noindent In \citep{Parkes1972}, based on structured interviews, it is observed that widowers have experienced disturbance of appetite and sleep, depression, restlessness during a period of 2 to 4 years after the bereavement. In \citep{Young1963}, it is shown that the mortality rate of the survived couple increases by 40\% during the first six months of bereavement and after decreases gradually to normal rate.
\par
\noindent Over the last few decades, several papers have been published which model future lifetimes of couples regarding to dependence, and pricing contingent coupled lives contracts. The papers show that there is a significant correlation between lifetimes of couples, causes substantial impact on relevant policies pricing. These models use variety of probabilistic tools in modelling the dependence between lifetimes. \citep{frees1996}, \citep{carrie2000}, \citep{luciano2008}, model the dependence of the time of deaths of coupled lives based on copulas.\citep{spre2008} uses Markov model to modelling the short term dependence between two remaining lifetimes and applied to a life annuity portfolio.
\par
\noindent Although the assumption of dependence between future lifetimes of couples makes the calculations straightforward and easy, the consequences can be unfair either for insurers or policyholders. Under independence assumption, the probability of joint survival of a couple with ages $x$ and $y$ is given by product of individual survival probability, \textit{i.e.} $_{n} p _{x,y} = _{n} p _{x} \times _{n} p _{y}$, where $ _{n} p _{x}$ is probability of survival of a person aged $x$ to age $x+n$ and $ _{n} p _{x,y}$ is probability of joint survival of a couples aged $x$ and $y$ to age $x+n$ and $y+n$. The individual survival probabilities can be calculated easily from actuarial life table.
\par
\noindent In order to include dependence between the future lifetimes of couples, multi-state models are widely used. A multi-state model is a model for a continuous-time stochastic process allowing the process to move among finite states. Multi-state models are widely used in demography, biostatistics and actuarial sciences.  When the stochastic process of the multi-state is a Markov process, the calculations are tractable. \citep{spre2008} uses Markov model and \citep{MinJiThesis} uses a semi-Markov model for modelling joint-life mortality. Disability insurance is modelled in a multi-state context in \citep{sver1965} and \citep{HassanZadeh.Disability.Saj}. Multi-state models also are used in modelling aging-process \citep{lin2007}.  \citep{ji2011} uses Markov and semi-Markov model to modelling dependence lifetimes for reverse mortgage terminations. For a thorough review of multi-state models and its applications in life insurance products see \citep{Act.Math.Dikson}.
\par
\noindent Modelling joint-life mortality presents challenges because of possibility of moving from joint-life status for different reasons into three other states, \textit{i.e.}, wife-dead husband-alive, husband-dead wife-alive and wife and husband both dead. Modelling a movement from an active joint-life status to the state when one of the couples has died needs special considerations. In this case, the broken-heart syndrome causes the rate of mortality of alive partner to move much higher than normal case.
In this paper, a new method for modelling joint-life mortality is presented based on Markov chain.
We also use a hypothetical health index called physiological age that representing the degree of aging in a human body. See \citep{lin2007} and referral inside it for additional information. \par
\noindent This paper is organized as following. In section \ref{s2} a brief introduction of phase-type distribution (\textit{PH}) is given. Section \ref{s3} describes the model and section \ref{s4} includes the actuarial calculations. A numerical example is presented in section \ref{s5} and finally we conclude this paper with conclusions in \ref{s6}.   \\

\section{Phase-type distribution} \label{s2}
Phase-type distributions recently have been received attentions by actuaries due to nice properties they own. Closed form expressions, interpretable parameters, being dense in all positive support distributions and ability to model complex systems make \textit{PH} distributions very attractive and applicable in actuarial context. In \citep{lin2007}, by imposing a physiological age process on the underlying Markov chain of the \textit{PH}, mortality rates is successfully modelled. In order to model disability, recovery and death, the same technique as of \citep{lin2007} was used in \citep{HassanZadeh.Disability.Saj} and all actuarial expressions are obtained in closed-forms. In \citep{Drekic}, in the Sparre Andersen renewal models with \textit{PH} distributed claims, the distribution of deficit at ruin is found to be  of \textit{PH}.\citep{asmussen2000} applies \textit{PH} distributions to risk theory. Credibility theory in context of \textit{PH} distributions has been developed by \citep{amin}, \citep{amin_thesis} and \citep{Zang2014}. See \citep{JuneCai.1}, \citep{JuneCai.2} and \citep{HassanZadeh.Bilodeau} for applications of multivariate \textit{PH} in actuarial science. \\
Consider $\{Z_t, t \geq 0\}$ a right continuous-time Markov process on the finite state space $\Gamma =\{0,1, 2,\dots, m, \Delta\}$ with initial probability vector $\boldsymbol{\alpha}$ and infinitesimal generator matrix $\boldsymbol{\Lambda}$. We assume that $\Delta$ is the only absorbent state. In this case the matrix $\boldsymbol{\Lambda}$ can be written as
\begin{eqnarray}
\left(
\begin{matrix}
\textbf{Q} & \textbf{q} \\
\textbf{0}^{'} & 0
\end{matrix} \right).
\end{eqnarray}
\noindent Where the matrix $\textbf{Q}=(q_{ij}), i,j =0,1,2,\dots,m$ is a sub-intensity matrix ( a square matrix $\textbf{B}=(b_{ij}), i,j=1,\dots,k$ is called a sub-intensity matrix if $b_{ii}\le 0; \,b_{ij} \ge 0$ for $i \neq j$, and $\sum_{j=1}^{k} b_{ij} \le  0$, with strick inequality for at least one $i$;  for $i,j=1,\dots,k$) and $\textbf{q}$ is the exit vector to the absorbent state $\Delta$ which equals $-\textbf{Q} \textbf{1}$. Without loss of generality, we presume that $\alpha_{\Delta}=0$, \textit{i.e.} $P(Z_{0} \in \{\Delta\})=0$. The initial probability vector over the transients elements is denoted by $\boldsymbol{\pi}$ such that $\boldsymbol{\alpha}= (\boldsymbol{\pi},0)$.

Let's define $T=\inf\{t; Z_t \in \Delta \}$. In this case $T$ is said to follow a \textit{PH} distribution with representation $(\boldsymbol{\pi}, \textbf{Q})$
The probability density function $f_T(t)$ and survival function, $S_T(t)$  are given as follow
\begin{eqnarray}
f_T(t) &=&\boldsymbol{\pi} e^{\textbf{Q}x} \textbf{q} \\
S_T(t) &=&\boldsymbol{\pi} e^{\textbf{Q}x} \textbf{1}
\end{eqnarray}
\noindent Where exponential of a square matrix $\textbf{A}$ is defined as $e^{\textbf{A}}=\sum_{k=0}^{\infty} \frac{\textbf{A}^k}{k !}$ and $\textbf{1}$ is a column vector of $1$s with a proper dimension. See \citep{neuts1981} for more details and proofs.\\
Bivariate phase type (\textit{BPH}) distributions, in a version that we are interested, are defined be \citep{Assaf}. Suppose that $\Gamma_1$ and $\Gamma_2$ are two nonempty stochastically closed subsets ($E$ is said to be stochastically closed if once $Z_t$ has entered $E$, it never leaves) of $\Gamma$ such that
$\Gamma_1 \bigcap \Gamma_2=\{\Delta\}$. Random variable $\textbf{T}=(T_1,T_2)$ is called a \textit{BPH} if $T_i=\inf\{t; Z_t \in \Gamma_i$\}, $i=1, 2$. It can be shown that the survival function of $\textbf{T}$ equals to
\begin{equation}\label{survival_bph_1}
S(t_1,t_2)=\left\{ \begin{array}{cc}
\boldsymbol{\pi} e^{\textbf{Q}t_1} \textbf{g}_1 e^{\textbf{Q}(t_2-t_1)} \textbf{g}_2 \textbf{1} & \textrm{if} \, \, t_2 \ge t_1 \ge 0 \\
\boldsymbol{\pi} e^{\textbf{Q}t_2} \textbf{g}_2 e^{\textbf{Q}(t_1-t_2)} \textbf{g}_1 \textbf{1} & \textrm{if} \, \, t_1\ge t_2 \ge 0
\end{array} \right.
\end{equation}
where $\textbf{g}_k$, $k=1, 2$, is an $(m+1) \times (m+1)$ diagonal matrix whose $i$th diagonal element is $1$ if $i \in \Gamma_i ^{c}$, $i=1, 2$ and $0$ otherwise.
After simple calculations, the joint probability density function of $\textbf{T}$ of the absolutely continuous component is given be the following
\begin{equation}\label{density_joint_1}
f(t_1,t_2)=\left\{ \begin{array}{cc}
\boldsymbol{\pi} e^{\textbf{Q}t_1} \textbf{G}_1 e^{\textbf{T}(t_2-t_1)} \textbf{Qg}_2 \textbf{1} & \textrm{if} \, \, t_2 > t_1 > 0 \\
\boldsymbol{\pi} e^{\textbf{Q}t_2} \textbf{G}_2 e^{\textbf{Q}(t_1-t_2)} \textbf{Qg}_1 \textbf{1} & \textrm{if} \, \, t_1\ > t_2 > 0
\end{array} \right.
\end{equation}
Where $\textbf{G}_i=\textbf{Q}\textbf{g}_i-\textbf{g}_i\textbf{Q}$, for $i=1,2$.

\noindent The singular component on $t_1=t_2$, is very useful for our case of application. In \citep{Assaf} and \citep{HZBil}, the following quantity is given for this quantity
\begin{equation} \label{eq_singualr_1}
P(T_1=T_2>t)= \boldsymbol{\pi} e^{\textbf{Q}t} \textbf{Q}^{-1} \textbf{g}_1 \textbf{g}_2 \textbf{Q} \textbf{1}.
\end{equation}

By setting $t=0$ in \eqref{eq_singualr_1}, one can get the the following
\[ P(T_1=T_2)= \boldsymbol{\pi} \textbf{Q}^{-1} \textbf{g}_1 \textbf{g}_2 \textbf{Q} \textbf{1}.\]
Define $ \Gamma_0 = \Gamma-(\Gamma_1 \bigcup \Gamma_2) $. Based on the definition for the \textit{BPH} by \citep{Assaf}, the sub-intensity matrix \textbf{Q} can be written as follows
\begin{equation} \label{Qde}
\textbf{Q}=
\begin{bmatrix}
\textbf{Q}_0 & \textbf{Q}_{01} & \textbf{Q}_{02} \\
\textbf{0}   & \textbf{Q}_{1}  & \textbf{0}      \\
\textbf{0}   & \textbf{0}      & \textbf{Q}_2  
\end{bmatrix}.
\end{equation}
\noindent In this case \eqref{density_joint_1}, \eqref{survival_bph_1} and \eqref{eq_singualr_1} will be simplified as noted in \citep{Assaf}. In the next section the proposed model in this paper is explained in details.

\section{The model} \label{s3} 
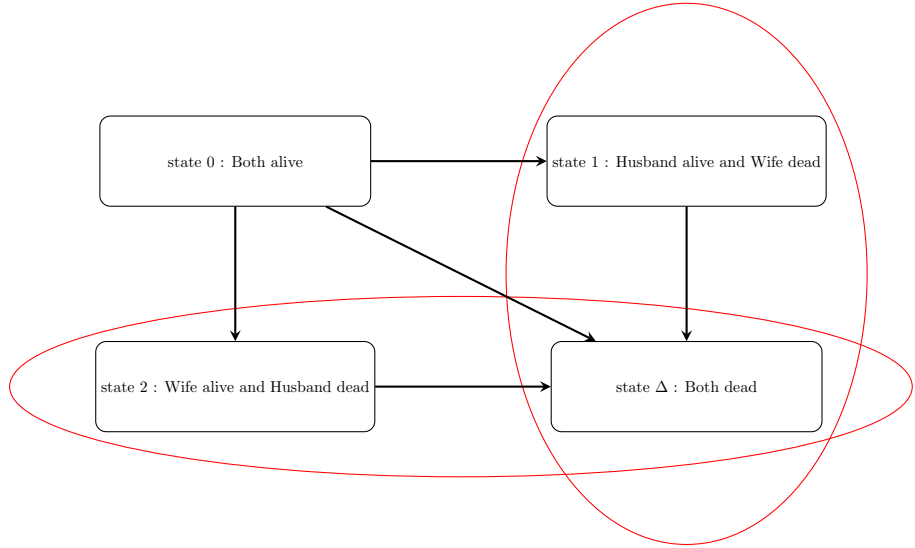
\begin{figure} [h!]
	
	\begin{center}
		
		\begin{tikzpicture}
		
		\tikzstyle{arrow1} = [thick,->,>=stealth,auto]	
		
		\tikzstyle{general} = [rectangle, rounded corners, minimum width=6cm, minimum height=2cm,text centered, draw=black,scale=.6]
		\tikzstyle{fg1} = [ellipse, minimum width=20cm, minimum height=4cm,text centered, draw=red,scale=.6]
		\tikzstyle{fg2} = [ellipse, minimum width=8cm, minimum height=12cm,text centered, draw=red,scale=.6]
		
		\node (bothalive)  [general]  {state 0 : Both alive};
		\node (mendead)  [general,right of=bothalive,node distance=10cm]  {state 1 : Husband alive and Wife dead};
		\node (womandead)  [general,below of=bothalive,node distance=5cm]  {state 2 : Wife alive and Husband dead};
		\node (bothdead)  [general,right of=womandead,node distance=10cm]  {state $\Delta$ : Both dead};
		\node(g1) at (3,-3) [fg1] {};
		\node(g2) at (6,-1.5) [fg2] {};
		
		\draw [arrow1] (bothalive) -- node {} (mendead) ;
		\draw [arrow1] (bothalive) -- node {} (womandead) ;
		\draw [arrow1] (mendead) -- node {} (bothdead) ;
		\draw [arrow1] (womandead) -- node {} (bothdead) ;
		\draw [arrow1] (bothalive) -- node {} (bothdead) ;
		
		\end{tikzpicture}
		
	\end{center}
	\caption{Multi-state model for joint and last survivor statuses .}
	\label{s3-f1}
\end{figure}

\noindent As it is mentioned in Section \ref{s1}, there are three main reasons which significantly affect the remaining lifetimes of a couple. In this section, we will develop the classic commos-shock model taking into account the dependence factors in a Markov chain environment. The model has four states, which explains survival status of a couple (see Figure \ref{s3-f1}).
\par
\noindent Let $\{Z_t, t\geq 0\}$ be a continuous-time Markov chain on finite state space $E=\{0,1,2,\Delta\}$ to represent the survival status of a couple at time $t$ in Figure \ref{s3-f1}, where the state $\Delta$ is absorbent and the rest are transient. In addition, we assume that in $t=0$ the husband (with real age $x$) and the wife (with real age $y$) are both alive, \textit{i.e.} $Z_0=0$. Suppose $T_x$ and $T_y$ to be two random variables that indicate the remaining lifetimes of the couple. We define
\begin{equation}
\begin{split}
T_{x}=\inf \{ t \geq 0 ; Z_t \in \Gamma_2\}, \\
T_{y}=\inf \{t \geq 0 ;  Z_t \in \Gamma_1\},
\end{split}
\end{equation}
where  $\Gamma_1=\{1,\Delta\}$ and $\Gamma_2=\{2,\Delta\}$ are two stochastically closed subsets of $ E $. Under this definition and the definition of \textit{BPH}, the random vector $(T_x,T_y)$ follows a \textit{BPH} distribution.
\par
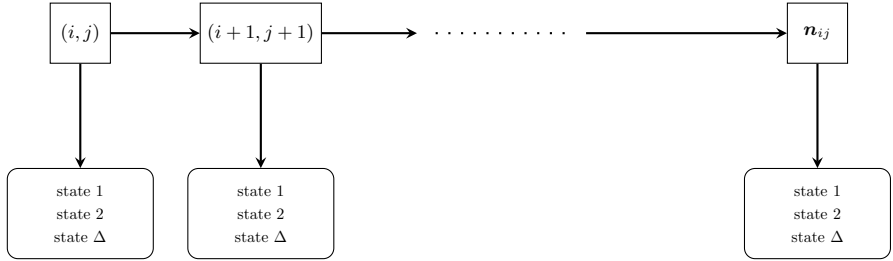
\begin{figure} [h!]
	
	\begin{center}
		
		\begin{tikzpicture}
		
		\tikzstyle{arrow1} = [thick,->,>=stealth,auto]	
		
		\tikzstyle{general} = [rectangle ,text centered, draw=black,scale=.8,minimum size=1cm]
		\tikzstyle{general2} = [rectangle, rounded corners, minimum width=3cm, minimum height=2cm,text centered, draw=black,scale=.6]
		
		\node (s1)  [general]  {\small$(i,j)$};
		\node (s2)  [general,right of=s1,node distance=3cm] {\small$(i+1,j+1)$};
		\node (s3) [right of=s2,node distance=2.2cm]  {};
		\node (s4) [right of=s3,node distance=2cm]  {};
		\node (s5) [general,right of=s4,node distance=4cm]  {\small $\boldsymbol{n}_{ij}$};
		\node (f1)  [general2,below of=s1,node distance=4cm,text width=3cm] {\text{state 1} \\ \text{state 2} \\ \text{state $\Delta$}};
		\node (f2)  [general2,below of=s2,node distance=4cm,text width=3cm] {\text{state 1} \\ \text{state 2} \\ \text{state $\Delta$}};
		\node (f3)  [general2,below of=s5,node distance=4cm,text width=3cm] {\text{state 1} \\ \text{state 2} \\ \text{state $\Delta$}};
		
		\draw [arrow1] (s1) -- node[above] {} (s2) ;
		\draw [arrow1] (s2) -- node[above] {} (s3) ;
		\draw [arrow1] (s4) -- node[above] {} (s5) ;
		\draw [arrow1] (s1) -- node[right] {} (f1) ;
		\draw [arrow1] (s2) -- node[right] {} (f2) ;
		\draw [arrow1] (s5) -- node[right] {} (f3) ;
		\draw [loosely dotted,thick] (s3) -- node[] {} (s4) ;
		
		\end{tikzpicture}
		
	\end{center}
	\caption{Decomposition of state $0$ in common shock model}
	\label{s3-f2}
\end{figure}

\noindent In order to reflect the causes of dependence in our model,  we need to impose some conditions that reflect both common life style and broken-heart syndrome effects in the model. To this end, in our model, each state is decomposed into sub-states. Decomposition of the state $0$ is presented as follows:
$$ \{ (i,j), (i+1,j+1), \dots,(n+i-max(i,j),n+j-max(i,j))\},$$ with a cardinality of  $d_0 : = n-max(i,j)+1$, where $i$ and $j$ denote the physiological age of the husband and the wife at issue of the insurance contract and $n$ is the maximum physiological age that a new born own. See Figure \ref{s3-f2}. ( we need more explanation and references about physiological age here ). In \cite{lin2007} it is shown that with $n=200$, a good fit for mortality rates of a newborn can be obtained. 
The direct transition from the state $0$ to the state $\Delta$ models the common shock effect. 
\par
 \noindent If the couple starts with the physiological age $(i,j)$, the process $Z_t$ either will run into the next physiological age $(i+1,j+1)$ with rate $\lambda$ or the process will move to the states $1$, $2$ or $\Delta$, as it is shown in Figure \ref{s3-f2}. 

In order to decompose the states $1$ and $2$ to appropriate sub-states, we have to consider the bereavement  effect on mortality. It's important to note that spousal death has a short-term effect on the survived partner. 

Thus, we use two sets of sub-states in the states $1$ and $2$ to reflect the bereavement effect. The first set includes sub-states that indicate the survived individual's physiological ages after spousal death. The second set represents the states after broken-heart syndrome effect period when the survived partner is in normal condition from mortality point of view.

\begin{figure} [h!]
	
	\begin{center}
		
		\begin{tikzpicture}
		
		\tikzstyle{arrow1} = [thick,->,>=stealth,auto]	
		
		\tikzstyle{general} = [circle ,text centered, draw=black,scale=.6,minimum size=2cm]
		\tikzstyle{general2} = [rectangle, rounded corners, minimum width=3cm, minimum height=2cm,text centered, draw=black,scale=.6]
		
		\node (s1)  [general]  {\small$j_{wf}$};
		\node (s2)  [general,right of=s1,node distance=3cm] {\small$(j+1)_{wf}$};
		\node (s3) [right of=s2,node distance=2.2cm]  {};
		\node (s4) [right of=s3,node distance=3cm]  {};
		\node (s5) [general,right of=s4,node distance=3cm]  {\small$n_{wf}$};
		\node (s11)  [general,below of=s1,node distance=3cm]  {\small$j_{f}$};
		\node (s22)  [general,right of=s11,node distance=3cm] {\small$(j+1)_{f}$};
		\node (s33) [right of=s22,node distance=2.2cm]  {};
		\node (s44) [right of=s33,node distance=3cm]  {};
		\node (s55) [general,right of=s44,node distance=3cm]  {\small$n_{f} $};
		\node (f1)  [above right of=s1,node distance=1.1cm] {};
		\node (f2)  [above right of=s2,node distance=1.1cm] {};
		\node (f5)  [above right of=s5,node distance=1.1cm] {};
		\node (f11)  [above right of=s11,node distance=1.1cm] {};
		\node (f22)  [above right of=s22,node distance=1.1cm] {};
		\node (f55)  [above right of=s55,node distance=1.1cm] {};
		\node (f111)  [above of=s1,node distance=1.5cm] {};
		\node (f222)  [above of=s2,node distance=1.5cm] {};
		\node (f555)  [above of =s5,node distance=1.5cm] {};
		
		\draw [arrow1] (s1) -- node[above] {} (s2) ;
		\draw [arrow1] (s2) -- node[above] {} (s3) ;
		\draw [arrow1] (s4) -- node[above] {} (s5) ;
		\draw [arrow1] (s1) -- node[right] {} (s22) ;
		\draw [arrow1] (s11) -- node[right] {} (s22) ;
		\draw [arrow1] (s22) -- node[right] {} (s33) ;
		\draw [arrow1] (s44) -- node[right] {} (s55) ;
		\draw [arrow1] (s1) -- node[above] {} (f1) ;
		\draw [arrow1] (s2) -- node[above] {} (f2) ;
		\draw [arrow1] (s5) -- node[above] {} (f5) ;
		\draw [arrow1] (s11) -- node[above] {} (f11) ;
		\draw [arrow1] (s22) -- node[above] {} (f22) ;
		\draw [arrow1] (s55) -- node[above] {} (f55) ;
		\draw [arrow1] (f111) -- node[above] {} (s1) ;
		\draw [arrow1] (f222) -- node[above] {} (s2) ;
		\draw [arrow1] (f555) -- node[above] {} (s5) ;
		\draw [arrow1] (s2) -- node[above] {} (s33) ;
		
		\draw [loosely dotted,thick] (s3) -- node[] {} (s4) ;
		\draw [loosely dotted,thick] (s33) -- node[] {} (s44) ;
		
		\end{tikzpicture}
		
	\end{center}
	\caption{Decomposition of state $2$ in common shock model}
	\label{s3-f3}
\end{figure}
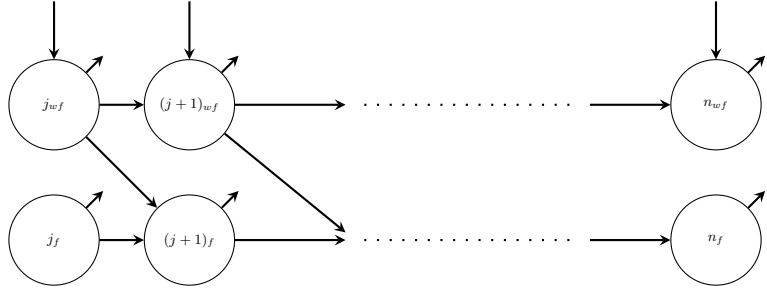

In the first phase of mortality, under the influence of bereavement effect, the rate of mortality is higher than the second one. We use subscripts $wm$ ($wf$) and $m$ ($f$) for physiological ages in bereavement (first phase) and after bereavement (second phase) of the husband (wife). In other words, symbols  $k_{wm}$ ($k_{wf}$) and $k_m$ ($k_f$) represent the husband (wife) physiological age $k$ under the influence of spousal death and after the broken-heart effect vanished, respectively. The result can be seen in Figure \ref{s3-f3} for decomposition of the state 2. As we see in Figure \ref{s3-f3}, the state 2 can be decomposed into $ 2d_2 $, where $d_2:= n-j+1$  states, as below:
$$ \{ j_{wf},(j+1)_{wf},\cdots,n_{wf},j_f,(j+1)_f,\cdots,n_f \} $$.

Where symbols
\begin{tikzpicture}

\tikzstyle{arrow1} = [thick,->,>=stealth,auto]	

\tikzstyle{general} = [circle, draw=black,scale=.6]

\node (s1)  [general]  {};
\node (f1)  [above right of=s1,node distance=.5cm] {};

\draw [arrow1] (s1) -- node[above] {} (f1) ;

\end{tikzpicture}
and
\begin{tikzpicture}

\tikzstyle{arrow1} = [thick,->,>=stealth,auto]	

\tikzstyle{general} = [circle, draw=black,scale=.6]
\node (s1)  [general]  {};
\node (f1)  [above of=s1,node distance=.5cm] {};
\draw [arrow1] (f1) -- node[above] {} (s1) ;
\end{tikzpicture}
indicate the process meets the $\Delta$ state and transition from state $0$ to state $2$, respectively. As it can be seen in Figure \ref{s3-f3}, after husband death, the process enters to state $2$ through the first phase of sub-states with $wf$ subscript, \textit{i.e.} $ Z_t \in \{ j_{wf},(j+1)_{wf},\cdots,n_{wf} \} $. In first phase the wife has further force of mortality, the cause of the broken-heart effect. After a while, provided that the wife is survived, the process moves to the second phase of sub-states with $f$ subscript, \textit{i.e.} $ Z_t \in \{ j_{f},(j+1)_{f},\cdots,n_{f} \} $.

\begin{figure} [h!]
	
	\begin{center}
		
		\begin{tikzpicture}
		
		\tikzstyle{arrow1} = [thick,->,>=stealth,auto]	
		
		\tikzstyle{general} = [circle ,text centered, draw=black,scale=.6,minimum size=2cm]
		\tikzstyle{general2} = [rectangle, rounded corners, minimum width=3cm, minimum height=2cm,text centered, draw=black,scale=.6]
		
		\node (s1)  [general]  {\small$i_{wm}$};
		\node (s2)  [general,right of=s1,node distance=3cm] {\small$(i+1)_{wm}$};
		\node (s3) [right of=s2,node distance=2.2cm]  {};
		\node (s4) [right of=s3,node distance=3cm]  {};
		\node (s5) [general,right of=s4,node distance=3cm]  {\small$n_{wm}$};
		\node (s11)  [general,below of=s1,node distance=3cm]  {\small$i_{m}$};
		\node (s22)  [general,right of=s11,node distance=3cm] {\small$(i+1)_{m}$};
		\node (s33) [right of=s22,node distance=2.2cm]  {};
		\node (s44) [right of=s33,node distance=3cm]  {};
		\node (s55) [general,right of=s44,node distance=3cm]  {\small$n_{m} $};
		\node (f1)  [above right of=s1,node distance=1.1cm] {};
		\node (f2)  [above right of=s2,node distance=1.1cm] {};
		\node (f5)  [above right of=s5,node distance=1.1cm] {};
		\node (f11)  [above right of=s11,node distance=1.1cm] {};
		\node (f22)  [above right of=s22,node distance=1.1cm] {};
		\node (f55)  [above right of=s55,node distance=1.1cm] {};
		\node (f111)  [above of=s1,node distance=1.5cm] {};
		\node (f222)  [above of=s2,node distance=1.5cm] {};
		\node (f555)  [above of =s5,node distance=1.5cm] {};
		
		\draw [arrow1] (s1) -- node[above] {} (s2) ;
		\draw [arrow1] (s2) -- node[above] {} (s3) ;
		\draw [arrow1] (s4) -- node[above] {} (s5) ;
		\draw [arrow1] (s1) -- node[right] {} (s22) ;
		\draw [arrow1] (s11) -- node[right] {} (s22) ;
		\draw [arrow1] (s22) -- node[right] {} (s33) ;
		\draw [arrow1] (s44) -- node[right] {} (s55) ;
		\draw [arrow1] (s1) -- node[above] {} (f1) ;
		\draw [arrow1] (s2) -- node[above] {} (f2) ;
		\draw [arrow1] (s5) -- node[above] {} (f5) ;
		\draw [arrow1] (s11) -- node[above] {} (f11) ;
		\draw [arrow1] (s22) -- node[above] {} (f22) ;
		\draw [arrow1] (s55) -- node[above] {} (f55) ;
		\draw [arrow1] (s2) -- node[above] {} (s33) ;
		\draw [arrow1] (f111) -- node[above] {} (s1) ;
		\draw [arrow1] (f222) -- node[above] {} (s2) ;
		\draw [arrow1] (f555) -- node[above] {} (s5) ;
		\draw [arrow1] (s2) -- node[above] {} (s33) ;
		
		\draw [loosely dotted,thick] (s3) -- node[] {} (s4) ;
		\draw [loosely dotted,thick] (s33) -- node[] {} (s44) ;
		
		\end{tikzpicture}
		
	\end{center}
	\caption{Decomposition of state $1$ in common shock model}
	\label{s3-f4}
\end{figure}
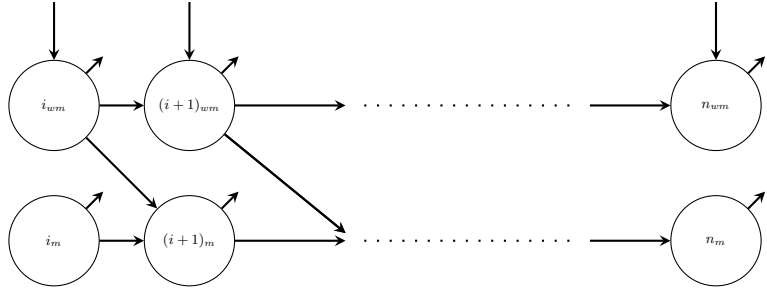

We use the second phase of states to show the wife aging process after recovery from the broken-heart syndrome. Decomposition of state $1$ is done in the same way, as for state 2. See Figure \ref{s3-f4}. This time with $ 2d_1 $, where $d_1:=n-i+1$,  states as follow:
\begin{align*}
\{ i_{wm},(i+1)_{wm},\cdots,n_{wm},i_m,(i+1)_m,\cdots,n_m \}
\end{align*}
\newpage
\begin{figure} [h!]
	\begin{center}
		
		\begin{tikzpicture}
		
		\tikzstyle{arrow1} = [thick,->,>=stealth,auto]	
		
		\tikzstyle{general} = [circle ,text centered, draw=black,scale=.6,minimum size=1.8cm]
		\tikzstyle{general1} = [rectangle ,text centered, draw=black,scale=.7,minimum size=1cm]
		
		\node (j1)  [general1]  {\small$(i,j)$};
		\node (j2)  [general1,below of=j1,node distance=2.5cm] {\small$(i+1,j+1)$};
		\node (j3)  [general1,below of=j2,node distance=2.5cm] {\small$(i+2,j+2)$};
		\node (j4) [below of=j3,node distance=2cm]  {};
		\node (j5) [below of=j4,node distance=2.5cm]  {};
		\node (j6) [general1,below of=j5,node distance=2.5cm]  {\small$\boldsymbol{n}_{ij}$};
		
		\node (fw1)  [general,right of=j1,node distance=4.5cm]  {\small$i_{wf}$};
		\node (fw2)  [general,below of=fw1,node distance=3cm] {\small$(i+1)_{wf}$};
		\node (fw3)  [general,below of=fw2,node distance=3cm]  {\small$(i+2)_{wf}$};
		\node (fw4) [below of=fw3,node distance=2.2cm]  {};
		\node (fw5) [below of=fw4,node distance=3cm]  {};
		\node (fw6) [general,below of=fw5,node distance=3cm]  {\small$ n_{wf} $};
		
		\node (f1)  [general,right of=fw1,node distance=3cm]  {\small$i_{f}$};
		\node (f2)  [general,below of=f1,node distance=3cm] {\small$(i+1)_{f}$};
		\node (f3)  [general,below of=f2,node distance=3cm] {\small$(i+2)_{f}$};
		\node (f4) [below of=f3,node distance=2.2cm]  {};
		\node (f5) [below of=f4,node distance=3cm]  {};
		\node (f6) [general,below of=f5,node distance=3cm]  {\small$ n_{f} $};
		
		\node (mw1)  [general,left of=j1,node distance=4.5cm]  {\small$i_{wm}$};
		\node (mw2)  [general,below of=mw1,node distance=3cm] {\small$(i+1)_{wm}$};
		\node (mw3)  [general,below of=mw2,node distance=3cm] {\small$(i+2)_{wm}$};
		\node (mw4) [below of=mw3,node distance=2.2cm]  {};
		\node (mw5) [below of=mw4,node distance=3cm]  {};
		\node (mw6) [general,below of=mw5,node distance=3cm]  {\small$ n_{wm} $};
		
		\node (m1)  [general,left of=mw1,node distance=3cm]  {\small$i_{m}$};
		\node (m2)  [general,below of=m1,node distance=3cm] {\small$(i+1)_{m}$};
		\node (m3)  [general,below of=m2,node distance=3cm] {\small$(i+2)_{m}$};
		\node (m4) [below of=m3,node distance=2.2cm]  {};
		\node (m5) [below of=m4,node distance=3cm]  {};
		\node (m6) [general,below of=m5,node distance=3cm]  {\small$ n_{m} $};
		
		\node (j01)  [above of=j1,node distance=1.5cm]  {};
		\node (j11)  [above right of=j1,node distance=1.2cm]  {};
		\node (j22)  [above right of=j2,node distance=1.2cm] {};
		\node (j33)  [above right of=j3,node distance=1.2cm] {};
		\node (j66) [above right of=j6,node distance=1.2cm]  {};
		
		\node (fw11)  [above right of=fw1,node distance=1.4cm]  {};
		\node (fw22)  [above right of=fw2,node distance=1.4cm] {};
		\node (fw33)  [above right of=fw3,node distance=1.4cm] {};
		\node (fw66) [above right of=fw6,node distance=1.4cm]  {};
		
		\node (f11)  [above right of=f1,node distance=1.4cm]  {};
		\node (f22)  [above right of=f2,node distance=1.4cm] {};
		\node (f33)  [above right of=f3,node distance=1.4cm] {};
		\node (f66) [above right of=f6,node distance=1.4cm]  {};
		
		\node (mw11)  [above left of=mw1,node distance=1.4cm]  {};
		\node (mw22)  [above left of=mw2,node distance=1.4cm] {};
		\node (mw33)  [above left of=mw3,node distance=1.4cm] {};
		\node (mw66) [above left of=mw6,node distance=1.4cm]  {};
		
		\node (m11)  [above left of=m1,node distance=1.4cm]  {};
		\node (m22)  [above left of=m2,node distance=1.4cm] {};
		\node (m33)  [above left of=m3,node distance=1.4cm] {};
		\node (m66) [above left of=m6,node distance=1.4cm] {}; 		  		 		 		 	 			 			
		
		\draw [arrow1] (j1) -- node[above] {} (j2) ;
		\draw [arrow1] (j2) -- node[above] {} (j3) ;
		\draw [arrow1] (j3) -- node[above] {} (j4) ;
		\draw [arrow1] (j5) -- node[above] {} (j6) ;
		
		\draw [arrow1] (fw1) -- node[above] {} (fw2) ;
		\draw [arrow1] (fw2) -- node[above] {} (fw3) ;
		\draw [arrow1] (fw3) -- node[above] {} (fw4) ;
		\draw [arrow1] (fw5) -- node[above] {} (fw6) ;
		
		\draw [arrow1] (f1) -- node[above] {} (f2) ;
		\draw [arrow1] (f2) -- node[above] {} (f3) ;
		\draw [arrow1] (f3) -- node[above] {} (f4) ;
		\draw [arrow1] (f5) -- node[above] {} (f6) ;
		
		\draw [arrow1] (mw1) -- node[above] {} (mw2) ;
		\draw [arrow1] (mw2) -- node[above] {} (mw3) ;
		\draw [arrow1] (mw3) -- node[above] {} (mw4) ;
		\draw [arrow1] (mw5) -- node[above] {} (mw6) ;
		
		\draw [arrow1] (m1) -- node[above] {} (m2) ;
		\draw [arrow1] (m2) -- node[above] {} (m3) ;
		\draw [arrow1] (m3) -- node[above] {} (m4) ;
		\draw [arrow1] (m5) -- node[above] {} (m6) ;
		
		\draw [arrow1] (j01) -- node[above] {} (j1) ;
		\draw [arrow1] (j1) -- node[above] {} (j11) ;
		\draw [arrow1] (j2) -- node[above] {} (j22) ;
		\draw [arrow1] (j3) -- node[above] {} (j33) ;
		\draw [arrow1] (j6) -- node[above] {} (j66) ;
		
		\draw [arrow1] (fw1) -- node[above] {} (fw11) ;
		\draw [arrow1] (fw2) -- node[above] {} (fw22) ;
		\draw [arrow1] (fw3) -- node[above] {} (fw33) ;
		\draw [arrow1] (fw6) -- node[above] {} (fw66) ;
		
		\draw [arrow1] (f1) -- node[above] {} (f11) ;
		\draw [arrow1] (f2) -- node[above] {} (f22) ;
		\draw [arrow1] (f3) -- node[above] {} (f33) ;
		\draw [arrow1] (f6) -- node[above] {} (f66) ;
		
		\draw [arrow1] (m1) -- node[above] {} (m11) ;
		\draw [arrow1] (m2) -- node[above] {} (m22) ;
		\draw [arrow1] (m3) -- node[above] {} (m33) ;
		\draw [arrow1] (m6) -- node[above] {} (m66) ;
		
		\draw [arrow1] (mw1) -- node[above] {} (mw11) ;
		\draw [arrow1] (mw2) -- node[above] {} (mw22) ;
		\draw [arrow1] (mw3) -- node[above] {} (mw33) ;
		\draw [arrow1] (mw6) -- node[above] {} (mw66) ;
		
		\draw [arrow1] (j1) -- node[above] {} (mw2) ;
		\draw [arrow1] (j2) -- node[above] {} (mw3) ;
		\draw [arrow1] (j1) -- node[above] {} (fw2) ;
		\draw [arrow1] (j2) -- node[above] {} (fw3) ;
		\draw [arrow1] (mw2) -- node[above] {} (m3) ;
		\draw [arrow1] (fw2) -- node[above] {} (f3) ;
		\draw [arrow1] (j3) -- node[above] {} (fw4) ;
		\draw [arrow1] (j3) -- node[above] {} (mw4) ;
		
		\draw [loosely dotted,thick] (j4) -- node[] {} (j5) ;
		\draw [loosely dotted,thick] (f4) -- node[] {} (f5) ;
		\draw [loosely dotted,thick] (m4) -- node[] {} (m5) ;
		\draw [loosely dotted,thick] (fw4) -- node[] {} (fw5) ;
		\draw [loosely dotted,thick] (mw4) -- node[] {} (mw5) ;

		\end{tikzpicture}
		
	\end{center}
	\caption{The \textit{BPH} model at a glance}
	\label{s3-f5}
\end{figure}
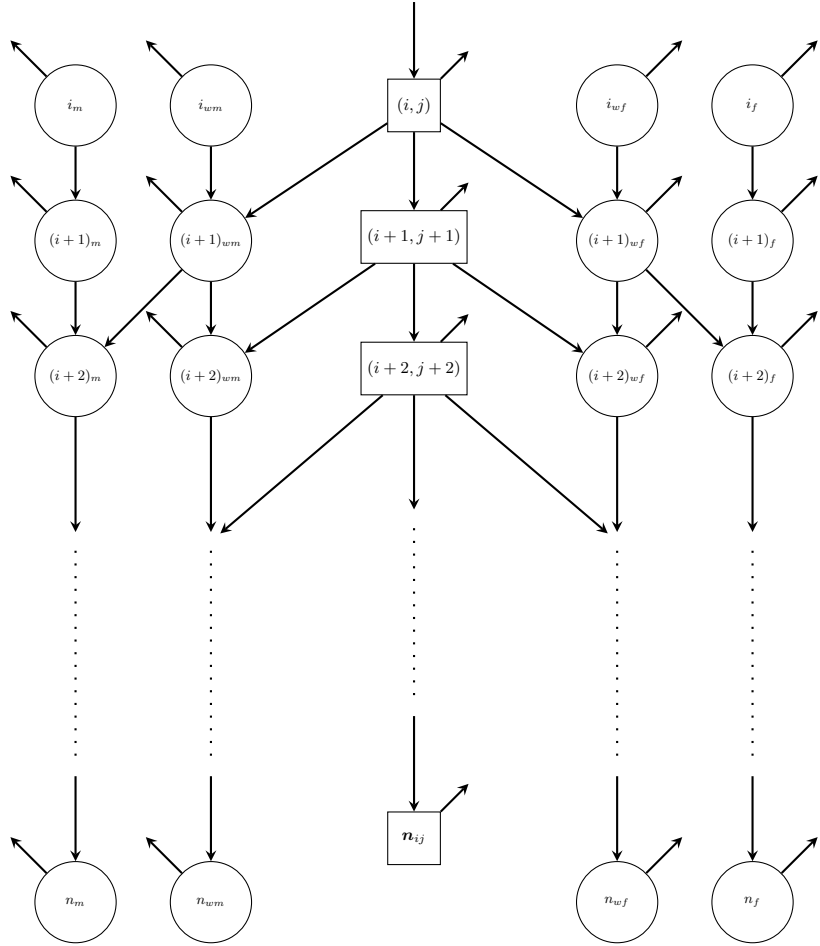
The entire model is depicted in Figure \ref{s3-f5}.
\par
Next step, after we have determined the structure of the model, we will explain the paramteres of the sub-intensity matrix $Q$ in the underlying Markov chain $\{Z_t, t\geq 0\}$.

In the following we will specify the elements of the sub-intensity matrix $\textbf{Q}$ in \eqref{Qde} by determining the elements of each sub-intensity matrix separately. We will use the rates given in \citep{lin2007} with a slight difference. In the joint status (state $0$), the rate of death for the joint status in physiological age $(i,j)$ is given by  
\begin{equation} \label{joint_mortality_rate} a_m + b_m i^{c_m} + a_f  + b_f j^{c_f}. \end{equation}
 
 The joint status   fails as soon as the first member dies and this explains the choice of the rate of mortality of joint status in \eqref{joint_mortality_rate}. The subscript $m$ ($f$) for the parameter in \eqref{joint_mortality_rate} is abbreviation for male (female).  The form and interpretation of these two intensities are as defined in \citep{lin2007}, but here the parameters are different for the husband and wife cases. Therefore the matrix $ \textbf{Q}_0 $ can be written as below.
\\ 
\begin{small}
\begin{align} \label{Q0}
\begin{dcases}
Q_0 (l,l) = -[\lambda _c + a_f + b_f (j+l-1)^{c_f} + a_m  + b_m (i+l-1)^{c_m} + \lambda]  ; & l < d_0 \\ 
Q_0 (l,l) = -[\lambda _c +  a_f + b_f (j+d_0-1)^{c_f}] ;  & l = d_0 , \,  i < j\\  
Q_0 (l,l) = -[\lambda _c + a_m  + b_m (i+d_0-1)^{c_m}]  ;  & l = d_0 ,  \,  i>j\\   
Q_0 (l,l+1)=\lambda ; &  l < d_0
\end{dcases}
\end{align}
\end{small}
\\

Other elements of $\textbf{Q}_0$ are zero. The parameter $ \lambda $ is used for rate of progression of aging of joint status of the couple. In other words, we use $ \lambda $ for transition rate from one pair physiological age to the next one. The parameter $\lambda_{c}$ is used for common shock event. 
The matrices $ \textbf{Q}_{01} $ and $ \textbf{Q}_{02} $, in \eqref{Qde} are  $d_0 \times 2 d_1$ and $d_0 \times 2 d_2$ matrices which contain the rates of a single death and $2d_1$ and $2d_2$ are dimensions of the matrices $\textbf{Q}_1$ and $\textbf{Q}_2$, respectively. The Non-zero elements of the matrix $\textbf{Q}_{01}$ and $\textbf{Q}_{02}$ can be written as follows.
\\
\begin{align} \label{Q02}
\begin{dcases}
Q_{01} (l,l+1)=a_f + b_f (j+l-1)^{c_f}  ; & l < d_0 \\ 
Q_{01} (l,l+1)=a_f + b_f (j+l-1)^{c_f}  ; &  l = d_0 , \,  j>i
\end{dcases}
\end{align}
\\
\begin{align}\label{Q01}
\begin{dcases}
Q_{02} (l,l+1)=a_m + b_m (i+l-1)^{c_m} ; & l < d_0 \\ 
Q_{02} (l,l+1)=a_m + b_m (i+l-1)^{c_m} ; &  l = d_0 , \, i>j
\end{dcases}
\end{align}
\\
Now it's  enough to determine the elements of sub-matrices $ \textbf{Q}_1 $ and $ \textbf{Q}_2 $. As it is mentioned before, we will use two different sets of mortality rates  to reflect the effect of the broken-heart syndrome after bereavement. The first set is for the short term high mortality rates after bereavement and the second set for after recovery. Therefore the matrix $\textbf{Q}_1$, is formulated as follows 
\begin{equation} \label{Qm}
\textbf{Q}_1=
\begin{bmatrix}
\textbf{Q}_{m1} & \textbf{Q}_{m12} \\
\textbf{0}      & \textbf{Q}_{m2}    
\end{bmatrix}.
\end{equation}
In order to magnify rates of mortality after breavement, we employ two parameters $\lambda_{wf}$ and $ \lambda_{wm} $ as multipliers of mortlaity rates of the wife and the husband, respectively. The effect of bereavement lasts for a while and then vanishes. The elements of the matrix $\textbf{Q}_{1}$ is given in the following.
\\
\begin{align}
\begin{dcases}
Q_{m1}(l,l+1) = \lambda _{in} ; & l<d_1 \\
Q_{m1}(l,l) =-[\lambda _{rm} +  \lambda _{in} + \lambda_{wm} (a_m+ b_m (i+l-1) ^{c_m})]; &  l<d_1 \\
Q_{m1}(l,l) =-\lambda_{wm} (a_m+ b_m (i+l-1) ^{c_m}); &  l=d_1 \\
\end{dcases},
\end{align}
\\
\begin{align}
Q_{m12}(l,l+1) =\lambda_{rm} ; \quad  l<d_1 ,
\end{align}
\\
and
\\
\begin{align}
\begin{dcases}
Q_{m2}(l,l+1) = \lambda _{in} ; & l<d_1 \\
Q_{m2}(l,l) =-[ \lambda _{in} +a_m+ b_m (i+l-1) ^{c_m}] ; &  l<d_1 \\
Q_{m2}(l,l) =-[ a_m+ b_m (i+l-1) ^{c_m}] ; &  l=d_1 \\
\end{dcases}.
\end{align}
\\
where $ \lambda_{in} $ is used for integrated measure of the deteriorating intensity of aging for a  widower. The matrix $ \boldsymbol{Q}_2 $ has the same structure as $ \boldsymbol{Q}_1 $, but the parameters $ a_m $, $ b_m $, $ c_m $, $ \lambda_{rm} $, $ \lambda_{wm} $  are replaced with $ a_f $, $ b_f $, $ c_f $, $ \lambda_{rf} $ and $ \lambda_{wf} $, respectively.  
\par
Hence, the structure of the proposed model  completely determined by the intensity matrix. In the next section, we will use the results obtained in this section to derive some essential actuarial quantities under the proposed model.

\section{Actuarial Present Value (APV) Calculations} \label{s4}
We assume that the process $\{Z_t, t\geq 0\}$ begin from state $ 0 $ with probability 1, \textit{i.e.} $ Pr[Z_0 \in \Gamma_0]=1 $.Therefore the initial probability vector can be written as $\boldsymbol{\pi} =(\boldsymbol{\pi}_0,\boldsymbol{0},\boldsymbol{0})$, where $ \boldsymbol{\pi}_0=(1,0,\cdots,0) $. Hence, the lifetimes of the couple $ (T_x,T_y) $ follow \textit{BPH} distribution with presentation $ (\boldsymbol{\pi},\boldsymbol{Q}) $. As mentioned in \citep{Assaf} the individual lifetimes will follow \textit{PH} distribution with representation $ (\boldsymbol{\pi}_x,\boldsymbol{Q}_x) $ and $ (\boldsymbol{\pi}_y,\boldsymbol{Q}_y) $, respectively, where 
\begin{align*}
\boldsymbol{\pi}_x =(\boldsymbol{\pi}_0,\textbf{0}), \quad \boldsymbol{\pi}_y =(\boldsymbol{\pi}_0,\textbf{0}).
\end{align*} 
and
\begin{align*}
\boldsymbol{Q_x}=\begin{bmatrix}
\boldsymbol{Q_0} & \boldsymbol{Q_{01}} & \\
\textbf{0}  & \boldsymbol{Q_1}
\end{bmatrix}
,
\quad
\boldsymbol{Q_y}=\begin{bmatrix}
\boldsymbol{Q_0} & \boldsymbol{Q_{02}} & \\
\textbf{0}  & \boldsymbol{Q_2}
\end{bmatrix}.
\end{align*}
\\
Also, it's easy to note that $\min(T_x,T_y)$ follows univariate \textit{PH} with represantation $(\boldsymbol{\pi}_0, \boldsymbol{Q}_0)$. 

Now we will take advantages of the properties of \textit{PH}(\textit{BPH}) distribution's  and of the underlying Markov process, $\{Z_t, t\geq 0\}$, to derive essential actuarial quantities. As we will see all the quantities have closed-form expressions. Some of these quantities are defined as follows.
\begin{itemize}
	\item
	${}_tP_{x:y}^{00}= Pr$[($ x $) and ($y$) are both alive in $t$ years.]
	\item
	${}_tP_{x}= Pr$[($ x $) is alive in $ t $ years.]
	\item
	${}_tP_{y}= Pr$[($y$) is alive in $ t $ years.]
	\item 
	${}_tP_{x:y}^{01}= Pr$[($y$) dies within $t$ years  and ($x$) is alive at $t$.]
	\item
	${}_tP_{x:y}^{02}= Pr$[($x$) dies within $t$ years  and ($y$) is alive at $t$.]
\end{itemize}
The $ {}_tP_{x:y}^{00} $ can be derived as following:
\begin{align*} 
{}_tP_{x:y}^{00} & = Pr(J_{t} \in \Gamma _0 | J_0 \in \Gamma_0 )= Pr(J_{t} \in \Gamma _0 ) & \\
& = Pr(T_x >t,T_y >t) &  \\
& =Pr[\min(T_x,T_y)>t]& \\
& = \boldsymbol{\pi}_0 exp\{\boldsymbol{Q}_0 t\} \boldsymbol{e}.
\end{align*}
Also, by using definition of ${}_tP_{x}$ we have: 
\begin{align*} 
{}_tP_x = Pr (T_x > t) = \boldsymbol{\pi}_x exp\{ \boldsymbol{Q}_x \}  e.
\end{align*}
${}_tP_{y}$ can be derived analogous to ${}_tP_{x}$. According to \citep{Act.Math.Dikson}, the probabilities ${}_tP_{x:y}^{01}$ and ${}_tP_{x:y}^{02}$ are given by:
\begin{align*}
{}_tP_{x:y}^{01} ={}_tP_{x}-{}_tP_{x:y}^{00}
\end{align*}
\begin{align*}
{}_tP_{x:y}^{02} ={}_tP_{y}-{}_tP_{x:y}^{00}
\end{align*}
Now, we will determine the actuarial present values (APV) of annuities and life insurane products in joint and last-survivor contracts. The definition of these quantities can be found in \citep{Act.Math.Dikson}. 

For the joint life annuity, APV is given by:
\begin{align*}
\bar{a}_{xy} & =  \int_{0}^{\infty} e^{-\delta t} {}_tP_{x:y}^{00} dt. & \\
& =   \int_{0}^{\infty} e^{-\delta t} \boldsymbol{\pi}_0  exp \{ \boldsymbol{Q}_0 t \} \boldsymbol{e} dt. & \\
& = \boldsymbol{\pi}_0 \int_{0}^{\infty} e^{-\delta t}   exp \{ \boldsymbol{Q}_0 t \}  dt \boldsymbol{e} & \\
& = \boldsymbol{\pi}_0 \int_{0}^{\infty}   exp \{ (\boldsymbol{Q}_0-\delta I) t \}  dt \boldsymbol{e} = \boldsymbol{\pi}_0 (\boldsymbol{Q}_0-\delta I) \boldsymbol{e}.
\end{align*}
Where $ \delta $ is the force of interest and $I$ is an identity matrix with an appropriate dimension. APV of a single life annuity issued for $ (x) $, can be derived as below.
\begin{align*}
\bar{a}_{x} & =  \int_{0}^{\infty} e^{-\delta t} {}_tP_{x} dt =   \int_{0}^{\infty} e^{-\delta t} \boldsymbol{\pi}_x exp \{ \boldsymbol{Q}_x t \} \boldsymbol{e} dt &\\
& = \boldsymbol{\pi}_x \int_{0}^{\infty} e^{-\delta t}   exp \{ \boldsymbol{Q}_x t \}  dt \boldsymbol{e} & \\
& = \boldsymbol{\pi}_x \int_{0}^{\infty}   exp \{ (\boldsymbol{Q}_x -\delta I) t \}  dt \boldsymbol{e} = \boldsymbol{\pi}_x (\boldsymbol{Q}_x -\delta I) \boldsymbol{e}.
\end{align*}
The APV of the last survivor annuity and the reversionary annuity can be determined by (see \cite{Act.Math.Dikson} for the definition):
\begin{align*}
\bar{a}_{\overline{xy}}=\bar{a}_{x}+\bar{a}_{y}-\bar{a}_{xy}, \qquad
\bar{a}_{x|y}=\bar{a}_{y}-\bar{a}_{xy}.
\end{align*}
APVs of the life insurance contracts  through the relationship between the APV of a life annuity and a life insurance contracts (see \cite{Act.Math.Dikson}). 
\begin{align*}
\bar{A} _{xy} = 1- \delta \bar{a}_{xy},
\end{align*}
\begin{align*}
\bar{A} _{x} = 1- \delta \bar{a}_{x},
\end{align*}
\begin{align*}
\bar{A} _{y} = 1- \delta \bar{a}_{y},
\end{align*}
and
\begin{align*}
\bar{A} _{\overline{xy}} = \bar{A} _{x}+\bar{A} _{y}-\bar{A} _{xy}.
\end{align*}

\section{A Numerical Example} \label{s5} 
As we have already seen, in the proposed model all the basic quantities that are attractive to actuaries have closed-form expressions. In this section, we will compute these quantities through the derived equations from the earlier section. The computations will show that the model reflects all the dependencies of lifetimes of a couple.
\\[.5cm]
\textbf{Example}: In this example we use results in \citep{lin2007} for setting some of the model parameters. We have assumed that the rate of mortality of male and female in the state $0$ are the same.  The parameters are given in Table \ref{ta1}.
\begin{table}[h!]
	\begin{tabular}{c  c  c c c c c} 
		$a_f$ & $b_f$ & $c_f$ & $a_m$ & $b_m$ & $c_m$ & $\lambda_{c}$ \\
        \hline
        \hline
		$9.0987e-04$ & $1.8872e-15$ & $6$ & $9.0987e-04$ & $1.8872e-15$  & $6.5$ & $.0002$  \\  
		$\lambda_{in}$ & $\lambda$ & $\lambda_{rf}$ & $\lambda_{rm}$ & $\lambda_{wf}$ & $\lambda_{wm}$ & $n$ \\
		\hline
		\hline
		$2.3707$ & $2.2$ & $5$ & $10$ & $4$ & $6$ & $200$  \\  
	\end{tabular}
	\caption{Parameters values in Example 1}
	\label{ta1}
\end{table}

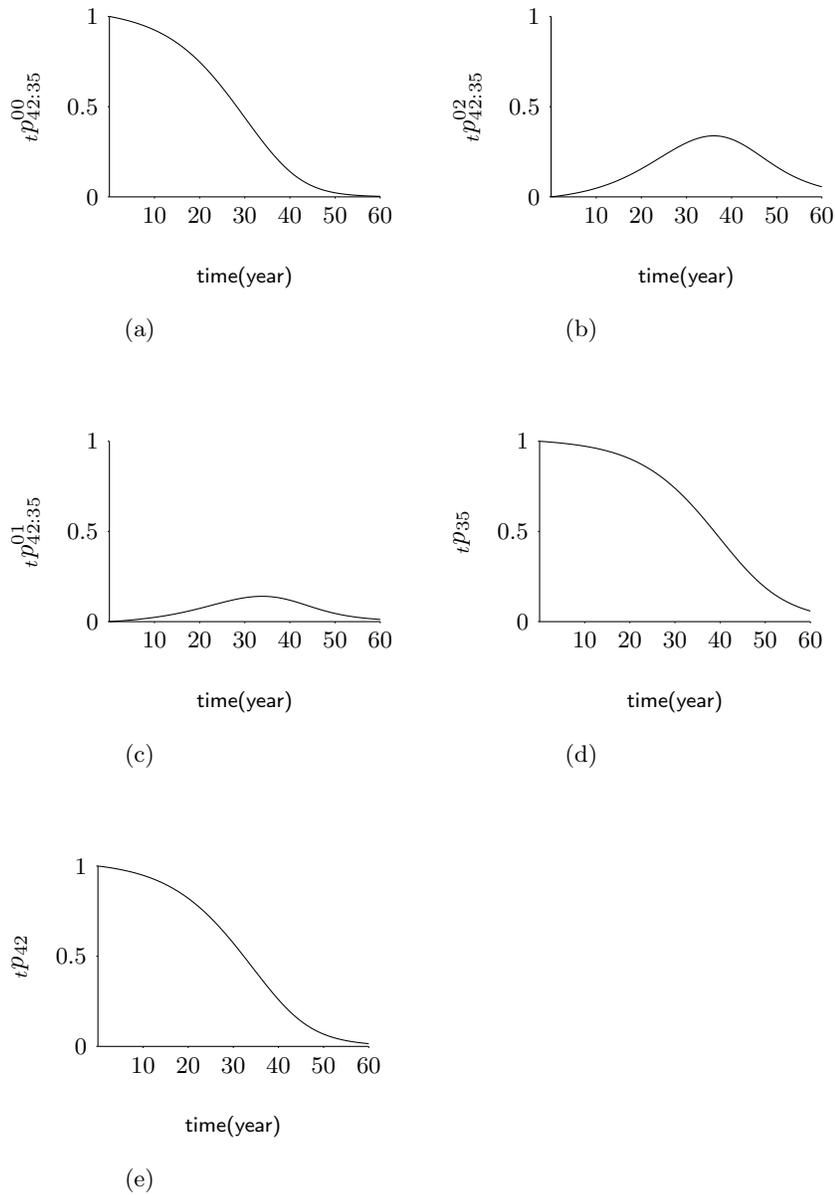
\begin{figure}[h!]
	\begin{subfigure}[b]{0.3\textwidth}
		\begin{tikzpicture}[y=4cm, x=.1cm,font=\sffamily,scale=.6]
		
		\draw (0,0) -- coordinate (x axis mid) (60,0);
		\draw (0,0) -- coordinate (y axis mid) (0,1);
		
		\foreach \x in {10,20,...,60}
		\draw (\x,0pt) -- (\x,-1pt)
		node[anchor=north] {\small{$\x$}};
		\foreach \y in {0,0.5,...,1}
		\draw (0pt,\y) -- (-1pt,\y) 
		node[anchor=east] {\small{$\y$}}; 
		
		\node[below=0.8cm] at (x axis mid) {\footnotesize{time(year)}};
		\node[rotate=90, above=0.8cm] at (y axis mid) {${}_tp^{00}_{42:35}$};
		
		\draw plot[] 
		file {p00.data};
		
		\end{tikzpicture}
		\caption{} 
		\label{s5f1a}
	\end{subfigure}
	\hspace{2cm}
	\begin{subfigure}[b]{0.3\textwidth}
		\begin{tikzpicture}[y=4cm, x=.1cm,font=\sffamily,scale=.6]
		
		\draw (0,0) -- coordinate (x axis mid) (60,0);
		\draw (0,0) -- coordinate (y axis mid) (0,1);
		
		\foreach \x in {10,20,...,60}
		\draw (\x,0pt) -- (\x,-1pt)
		node[anchor=north] {\small{$\x$}};
		\foreach \y in {0,0.5,...,1}
		\draw (0pt,\y) -- (-1pt,\y) 
		node[anchor=east] {\small{$\y$}}; 
		
		\node[below=0.8cm] at (x axis mid) {\footnotesize{time(year)}};
		\node[rotate=90, above=0.8cm] at (y axis mid) {${}_tp^{02}_{42:35}$};
		
		\draw plot[] 
		file {p01.data};
		
		\end{tikzpicture}
		\caption{} 
		\label{s5f1b}
	\end{subfigure}
	\\[1cm]
	\begin{subfigure}[b]{0.3\textwidth}
		\begin{tikzpicture}[y=4cm, x=.1cm,font=\sffamily,scale=.6]
		
		\draw (0,0) -- coordinate (x axis mid) (60,0);
		\draw (0,0) -- coordinate (y axis mid) (0,1);
		
		\foreach \x in {10,20,...,60}
		\draw (\x,0pt) -- (\x,-1pt)
		node[anchor=north] {\small{$\x$}};
		\foreach \y in {0,0.5,...,1}
		\draw (0pt,\y) -- (-1pt,\y) 
		node[anchor=east] {\small{$\y$}}; 
		
		\node[below=0.8cm] at (x axis mid) {\footnotesize{time(year)}};
		\node[rotate=90, above=0.8cm] at (y axis mid) {${}_tp^{01}_{42:35}$};
		
		\draw plot[] 
		file {p02.data};
		
		\end{tikzpicture}
		\caption{} 
		\label{s5f1c}
	\end{subfigure}
	\hspace{2cm}
	\begin{subfigure}[b]{0.3\textwidth}
		\begin{tikzpicture}[y=4cm, x=.1cm,font=\sffamily,scale=.6]
		
		\draw (0,0) -- coordinate (x axis mid) (60,0);
		\draw (0,0) -- coordinate (y axis mid) (0,1);
		
		\foreach \x in {10,20,...,60}
		\draw (\x,0pt) -- (\x,-1pt)
		node[anchor=north] {\small{$\x$}};
		\foreach \y in {0,0.5,...,1}
		\draw (0pt,\y) -- (-1pt,\y) 
		node[anchor=east] {\small{$\y$}}; 
		
		\node[below=0.8cm] at (x axis mid) {\footnotesize{time(year)}};
		\node[rotate=90, above=0.8cm] at (y axis mid) {${}_tp_{35}$};
		
		\draw plot[] 
		file {psw.data};
		\end{tikzpicture}
		\caption{} 
		\label{s5f1d}
	\end{subfigure}
	\\[1cm]
	\begin{subfigure}[b]{0.3\textwidth}
		\begin{tikzpicture}[y=4cm, x=.1cm,font=\sffamily,scale=.6]
		
		\draw (0,0) -- coordinate (x axis mid) (60,0);
		\draw (0,0) -- coordinate (y axis mid) (0,1);
		
		\foreach \x in {10,20,...,60}
		\draw (\x,0pt) -- (\x,-1pt)
		node[anchor=north] {\small{$\x$}};
		\foreach \y in {0,0.5,...,1}
		\draw (0pt,\y) -- (-1pt,\y) 
		node[anchor=east] {\small{$\y$}}; 
		
		\node[below=0.8cm] at (x axis mid) {\footnotesize{time(year)}};
		\node[rotate=90, above=0.8cm] at (y axis mid) {${}_tp_{42}$};
		
		\draw plot[] 
		file {psm.data};
		
		\end{tikzpicture}
		\caption{} 
		\label{s5f1e}
	\end{subfigure}
	\caption{The probabilities plots for the husband and the wife with real ages 42 and 35, respectively.}
	\label{s5f1}
\end{figure}

We assume that the couple are of real ages 42 (husband) and 35 (wife) in this example. The physiological age at issue, \textit{i.e.} $i$ and $j$ are assumed to be the mean of conditional expected value of the physiological age given the real age. The conditional probability vector of physiological given real age $x$ is given in the following:
\begin{equation*}
Pr[\textrm{physiological age}| \textrm{real age} = x]=\frac{\boldsymbol{\pi}e^{\textbf{Q}x}}{\boldsymbol{\pi}e^{\textbf{Q}x}\textbf{1}}
\end{equation*}  Figure \ref{s5f1} shows some quantities of interest. As stated in sub-Figures \ref{s5f1a}, \ref{s5f1d} and \ref{s5f1e}, these probabilities are decreasing functions of time. The remaining probabilities ${}_tp^{01}_{42:35}$ and ${}_tp^{02}_{42:35}$ in sub-Figures \ref{s5f1b} an \ref{s5f1c} increase at first and after around 38 years decrease to zero. The increase part is obvious as probability of death is an increasing function of age. The decrease part is interpretable as after 38 years the probability of death for the wife increases as well. The same interpretation is true for the \ref{s5f1c}. However, since the age of the wife is less than the age of the husband, the probability has the same shape as in \ref{s5f1b} but lesser than it.
\par
Other important quantities that discussed in latter section were APVs of annuities and life insurances. We can find the computed values of three different annuity contracts with different interest rates in Table \ref{ta2}.
\\
\begin{table}[h!]
	\centering
	\begin{tabular}{c | c  c  c } 
		Interest rate & $\bar{a}_{\overline{42:35}}$ & $\bar{a}_{42:35}$ & $\bar{a}_{35}$ \\
		\hline
		\hline
		5\% & $17.4444$ & $14.2534$ & $16.4525$  \\  
		10\% & $10.1519$ & $9.1281$ & $9.8199$  \\ 
		15\% & $7.0833$ & $6.6433$ & $6.9370$  \\
	\end{tabular}
	\caption{APVs of annuities contracts}
	\label{ta2}
\end{table}
\\
As shown in Table \ref{ta2}, since the payments is stopped when the first one dies, $\bar{a}_{35:42}$ has the smallest amount compared to the last surviver and single one in Table \ref{ta2}. This direction is opposite in the case of the corresponding life insurance contracts (see Table \ref{ta3}).
\\
\vspace{5cm}
\begin{table} [h!]
	\centering
	\begin{tabular}{c | c  c  c } 
		Interest rate  & $\bar{A}_{\overline{42:35}}$ & $\bar{A}_{42:35}$ & $\bar{A}_{35}$ \\
		\hline
		\hline
		5\% & $0.1489$ & $0.3046$ & $0.1973$  \\  
		10\% & $0.0324$ & $0.1300$ & $0.0641$  \\ 
		15\% & $0.0100$ & $0.0715$ & $0.0305$  \\
	\end{tabular}
	\caption{APVs of life insurance contracts}
	\label{ta3}
\end{table}
In the next step, we use conditional force of mortality, \textit{i.e.} $\mu_{T_x|T_y}$ and $\mu_{T_y|T_x}$ to evaluate the model in reflection of the broken-heart  syndrome effect. In \textit{BPH} distributions, conditional force of mortality has closed-form expression as mentioned in \citep{Assaf}.  $\mu_{T_x|T_y}(t_x|t_y)$ equals to

\begin{equation}
\frac{\boldsymbol{\pi}_0 e^{\boldsymbol{Q_0}\,t_y}\boldsymbol{Q_{01}}\,e^{\boldsymbol{Q_1}(t_x-t_y)}\boldsymbol{Q_1} \boldsymbol{1}}{\boldsymbol{\pi}_0 e^{\boldsymbol{Q_0}\,t_y}\boldsymbol{Q_{01}}\,e^{\boldsymbol{Q_1}(t_x-t_y)} \boldsymbol{1}}.
\end{equation}
A similar formula is available for  $\mu_{T_y|T_x}$

\begin{figure}[h!]
	\centering
	\begin{tikzpicture}[y=80cm, x=.5cm,font=\sffamily]
	
	\draw (19,0) -- coordinate (x axis mid) (30,0);
	\draw (19,0) -- coordinate (y axis mid) (19,.06);
	
	\foreach \x in {19,25,30}
	\draw (\x,0pt) -- (\x,-1pt)
	node[anchor=north] {\small{$\x$}};
	\foreach \y in {0,0.03,.06}
	\draw (19,\y) -- (18.9,\y) 
	node[anchor=east] {\small{$\y$}}; 
	
	\node[below=0.8cm] at (x axis mid) {\footnotesize{time(year)}};
	\node[rotate=90, above=0.8cm] at (y axis mid) {Conditional force of mortality};
	
	\draw[loosely dashed] plot[] 
	file {mufcon.data};
	
	\draw[] plot[] 
	file {mumcon.data};
	
	\begin{scope} 
	\draw[] (22,.05) -- (22.8,.05)
	node[right]{$ \mu_{T_x|T_y=20} $};
	\draw[loosely dashed] (22,.045) -- (22.8,.045)
	node[right]{$ \mu_{T_y|T_x=20} $};
	\end{scope}
	
	\end{tikzpicture}
	\caption{Conditional force of mortality}
	\label{s5f3}
\end{figure}
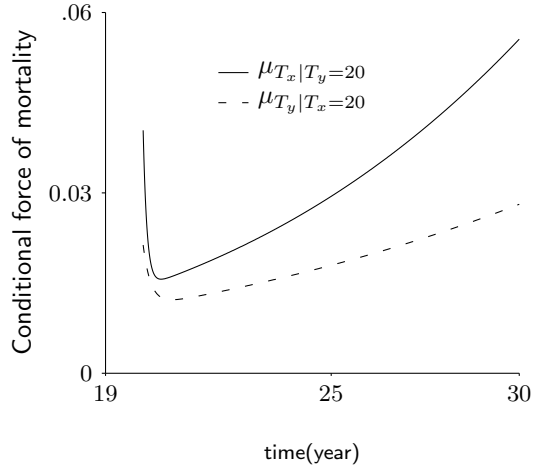
As it is seen from \ref{s5f3}, given that one of the couples dies after 20 years, the rate of mortality for the survived couple is high after the death of his/her couple. The broken-heart syndrom effect for the husband is stronger than for the wife. Since age of the man is higher than the wife, the total rate of mortality of the husband is higher than wife's. 
\section{ Conclusions and future of the research} \label{s6}
In this paper, we have presented a new model for modelling the future lifetime of a couple. Based on this model, the future lifetime follows a bivariate phase-type distribution. The model reflects dependence between future lifetimes of the wife and the husband. Some actuarial quantities are obtained. The work in this article can be developed by estimation of the parameters. The EM algorithm developed by (Asmussen 1996) is not appropriate as the E-step will be very slow. Another interesting topic can be studying dependence structure of the proposed model. Based of our simulation the correlation coefficient of random variable varies in a range of $[-\frac{1}{3}, 1]$. However, we could not prove it in theory.  
\newpage
\bibliographystyle{abbrvnat}

\bibliography{bibliography}

\begin{thebibliography}{23}
\providecommand{\natexlab}[1]{#1}
\providecommand{\url}[1]{\texttt{#1}}
\expandafter\ifx\csname urlstyle\endcsname\relax
  \providecommand{\doi}[1]{doi: #1}\else
  \providecommand{\doi}{doi: \begingroup \urlstyle{rm}\Url}\fi

\bibitem[Asmussen(2000)]{asmussen2000}
S.~Asmussen.
\newblock \emph{Ruin probabilities}, volume~2 of \emph{Advance series on
  statistical science \& applied probability}.
\newblock World Scientific, River Edge, N.J., 2000.

\bibitem[Assaf et~al.(1984)Assaf, Langberg, Savits, and Shaked]{Assaf}
D.~Assaf, N.~A. Langberg, T.~H. Savits, and M.~Shaked.
\newblock Multivariate phase-type distributions.
\newblock \emph{Operations Research}, pages 688--702, 1984.

\bibitem[Cai and Li(2005{\natexlab{a}})]{JuneCai.1}
J.~Cai and H.~Li.
\newblock Multivariate risk model of phase type.
\newblock \emph{Insurance: Mathematics and Economics}, pages 137--152,
  2005{\natexlab{a}}.

\bibitem[Cai and Li(2005{\natexlab{b}})]{JuneCai.2}
J.~Cai and H.~Li.
\newblock Conditional tail expectations for multivariate phase-type
  distributions.
\newblock \emph{Journal of Applied Probability}, pages 810--825,
  2005{\natexlab{b}}.

\bibitem[Carriere(2000)]{carrie2000}
J.~F. Carriere.
\newblock Bivariate survival models for coupled lives.
\newblock \emph{Scandinavian Actuarial Journal}, pages 17--32, 2000.

\bibitem[Dickson et~al.(2013)Dickson, Hardy, and Waters]{Act.Math.Dikson}
D.~C.~M. Dickson, M.~Hardy, and H.~R. Waters.
\newblock \emph{Actuarial Mathematics for Life Contingent Risks}.
\newblock Cambridge, 2013.

\bibitem[Drekic et~al.(2004)Drekic, Dickson, Stanford, and Willmot]{Drekic}
S.~Drekic, D.~C. Dickson, D.~A. Stanford, and G.~E. Willmot.
\newblock On the distribution of the deficit at ruin when claims are
  phase-type.
\newblock \emph{Scandinavian Actuarial Journal}, pages 105--120, 2004.

\bibitem[Frees et~al.(1996)Frees, Carriere, and Valdez]{frees1996}
E.~W. Frees, J.~Carriere, and E.~Valdez.
\newblock Annuity valuation with dependent mortality.
\newblock \emph{Journal of Risk and Insurance}, pages 229--261, 1996.

\bibitem[Hassan~Zadeh(2009)]{amin_thesis}
A.~Hassan~Zadeh.
\newblock \emph{Actuarial applications of multivariate phase-type
  distributions: Model calibration and credibility}.
\newblock PhD thesis, U. de Montreal, 2009.

\bibitem[Hassan-Zadeh and Bilodeau(2013)]{HZBil}
A.~Hassan-Zadeh and M.~Bilodeau.
\newblock Fitting bivariate distributions with phase-type distributions.
\newblock \emph{Scandinavian Actuarial Journal}, 2013:\penalty0 241--262, 2013.

\bibitem[Hassan-Zadeh and Stanford(2016)]{amin}
A.~Hassan-Zadeh and D.~A. Stanford.
\newblock Bayesian and b\"{u}hlmann credibility for phase-type distributions
  with a univariate risk parameter.
\newblock \emph{Scandinavian Actuarial Journal}, pages 338--355, 2016.

\bibitem[Hassan~Zadeh et~al.(2014)Hassan~Zadeh, Jones, and
  Stanford]{HassanZadeh.Disability.Saj}
A.~Hassan~Zadeh, B.~L. Jones, and D.~A. Stanford.
\newblock The use of phase-type models for disability insurance calculations.
\newblock \emph{Scandinavian Actuarial Journal}, pages 714--728, 2014.

\bibitem[Ji(2011)]{MinJiThesis}
M.~Ji.
\newblock \emph{Markovian Approaches to Joint-life Mortality with Applications
  in Risk Management}.
\newblock PhD thesis, U. of Waterloo, 2011.

\bibitem[Ji et~al.(2011)Ji, Hardy, and Li]{ji2011}
M.~Ji, M.~Hardy, and J.~S.~H. Li.
\newblock Markovian approaches to joint-life mortality.
\newblock \emph{North American Actuarial Journal}, pages 357--376, 2011.

\bibitem[Lin and Liu(2007)]{lin2007}
X.~S. Lin and X.~Liu.
\newblock Markov aging process and phase-type law of mortality.
\newblock \emph{North American Actuarial Journal}, 2007.

\bibitem[Luciano et~al.(2008)Luciano, Spreeuw, and Vigna]{luciano2008}
E.~Luciano, J.~Spreeuw, and E.~Vigna.
\newblock Modelling stochastic mortality for dependent lives.
\newblock \emph{Insurance: Mathematics and Economics}, pages 234--244, 2008.

\bibitem[Neuts(1981)]{neuts1981}
M.~Neuts.
\newblock \emph{Matrix-geometric solutions in stochastic models}, volume~2 of
  \emph{John Hoplins Series in the Mathematical Sciences.}
\newblock Johns Hopkins University Press, Baltimore, Md, 1981.

\bibitem[Parkes and Brown(1972)]{Parkes1972}
M.~C. Parkes and R.~J. Brown.
\newblock Health after bereavement: A controlled study of young boston widows
  and widowers.
\newblock \emph{Psychosomatic Medicine}, 1972.

\bibitem[Spreeuw(2008)]{spre2008}
.~W.~X. Spreeuw, J.
\newblock Modelling the short-term dependence between two remaining lifetimes.
\newblock \emph{Cass Business School Discussion Paper}, 2008.

\bibitem[Sverdrup(1965)]{sver1965}
E.~Sverdrup.
\newblock Estimates and test procedures in connection with stochastic models
  for deaths, recoveries and transfers between different states of health.
\newblock \emph{Scandinavian Actuarial Journal}, 1965.

\bibitem[Young et~al.(1963)Young, Benjamin, and Wallis]{Young1963}
M.~Young, B.~Benjamin, and C.~Wallis.
\newblock The mortality of widowers.
\newblock \emph{Lancet}, pages 454--6, 1963.

\bibitem[Zadeh and Bilodeau(2013)]{HassanZadeh.Bilodeau}
A.~H. Zadeh and M.~Bilodeau.
\newblock Fitting bivariate losses with phase-type distributions.
\newblock \emph{Scandinavian Actuarial Journal}, pages 241--262, 2013.

\bibitem[Zang(2013)]{Zang2014}
Y.~Zang.
\newblock \emph{Credibility with Phase--type Distributions}.
\newblock PhD thesis, U. Western Ontario, 2013.

\end{thebibliography}

\end{document}